\documentclass[preprint,amsmath,amssymb,aps,showkeys,showpacs]{revtex4}
\usepackage[english]{babel}
\usepackage{graphicx}
\usepackage{graphics}
\usepackage{amsmath}
\usepackage{dcolumn}
\usepackage{amssymb}
\usepackage{bm}

\begin{document}
\title{Lorentz symmetry breaking effects on relativistic EPR correlations}
\author{H. Belich} 
\email{belichjr@gmail.com}
\affiliation{Departamento de F\'isica e Qu\'imica, Universidade Federal do Esp\'irito Santo, Av. Fernando Ferrari, 514, Goiabeiras, 29060-900, Vit\'oria, ES, Brazil.}

\author{C. Furtado}
\email{furtado@fisica.ufpb.br}
\affiliation{Departamento de F\'isica, Universidade Federal da Para\'iba, Caixa Postal 5008, 58051-970, Jo\~ao Pessoa, PB, Brazil.}

\author{K. Bakke}
\email{kbakke@fisica.ufpb.br}
\affiliation{Departamento de F\'isica, Universidade Federal da Para\'iba, Caixa Postal 5008, 58051-970, Jo\~ao Pessoa, PB, Brazil.}

\begin{abstract}
Lorentz symmetry breaking effects on relativistic EPR (Einstein-Podolsky-Rosen) correlations are discussed. From the modified Maxwell theory coupled to gravity, we establish a possible scenario of the Lorentz symmetry violation and write an effective metric for the Minkowski spacetime. Then, we obtain the Wigner rotation angle via the Fermi-Walker transport of spinors and consider the WKB ((Wentzel-Kramers-Brillouin) approximation in order to study the influence of Lorentz symmetry breaking effects on the relativistic EPR correlations. 
\end{abstract}

\maketitle

\section{Introduction}

In special relativity, the Wigner rotation corresponds to the product of two Lorentz boots in different directions which gives rise to a boost preceded or followed by a rotation \cite{wigner,weinberg2}. Besides, the Wigner rotation is characterized by leaving the 4-momentum of the particle unchanged and making a precession of the spins in the rest frame of the particles. One effect associated with the Wigner rotation is the Thomas precession \cite{misner,thomas}. Another effect associated with the Wigner rotation is the precession of spins of the relativistic Einstein-Podolsky-Rosen (EPR) correlation \cite{epr} with respect to the initial configuration of spins due to the action of Lorentz transformations. This precession of spins yields an apparent deterioration of the initial correlations between the spins and decreases the degree of violation of the Bell inequality. In Refs. \cite{bell1,bell2,bell3,bell4,ahn}, it is shown that there exists decrease in the degree of the Bell inequality is yielded by the relativistic motion of the particle in the Minkowski spacetime. On the other hand, in curved spacetime, the decrease in the degree of the Bell inequality is yielded by the relativistic motion of the particles, the gravitational field and the position of the observers  \cite{bell5,tu2,bcf,bcf2,bcf3}. Other interesting studies of quantum entanglement in curved spacetime have been made in Refs. \cite{meh,meh2,meh3,meh4,meh5,meh6}.

A geometric approach was proposed by Borzeszkowski and Mensky \cite{repr} in order to study the relativistic EPR correlations in the presence of a gravitational field by applying the parallel transport along the world lines of the particles. However, Terashima and Ueda \cite{tu2} showed that by taking into account the accelerated motion of the particle and the gravitational field, thus, the parallel transport cannot yield the perfect direction of the relativistic EPR correlations. From this perspective, a geometric approach based on the Fermi-Walker transport has been proposed in Ref. \cite{bcf2} in order to obtain the Wigner rotation angle and the precession of the spins of a relativistic EPR correlation.

In this paper, we discuss the Lorentz symmetry breaking effects on relativistic EPR correlations. We start by introducing the description of fermions in curved spacetime in the presence of Lorentz symmetry breaking effects. The interest in studying the violation of the Lorentz symmetry, for instance, comes from the origin of electron electric dipole moment in which is not explained by the Standard Model of particle physics. At present days, just experimental upper bounds have been established \cite{revmod}. From this perspective, a necessity of investigating the physics beyond the Standard Model has arisen. A possible way of dealing with a scenario beyond the Standard Model is the extension of the mechanism for spontaneous symmetry breaking through vector or tensor fields, which implies that the Lorentz symmetry is violated. The seminal work made by Kosteleck\'{y} and Samuel \cite{extra3} in the string theory, where it is shown that the Lorentz symmetry is violated through a spontaneous symmetry breaking mechanism triggered by the appearance of nonvanishing vacuum expectation values of nontrivial Lorentz tensors, is considered to be the starting point for building several models that deal with the violation of the Lorentz symmetry. Such models are considered to be effective theories whose analysis of the phenomenological aspect at low energies may provide information and impose restrictions on the fundamental theory in which they stem from. In particular, a geometrical approach to investigating the effects of the violation of the Lorentz symmetry on photons was proposed in Refs. \cite{curv2,curv3}, where the Lagrangian of the modified Maxwell theory coupled to gravity is written in terms of an effective metric tensor. Therefore, our first objective in this work is to extend the geometrical approach proposed in Refs. \cite{curv2,curv3} to a fermionic field by modifying the Minkowski spacetime. Then, by establishing a possible scenario of the Lorentz symmetry violation, we wish to investigate the effects of the Lorentz violation on a gedanken experiment with relativistic entangled particles. From a classical mechanical point of view, we obtain the Wigner rotation angle via the Fermi-Walker transport of spinors. Further, from a quantum mechanical point of view, we consider the WKB approximation \cite{griff,anandan} and study the influence of Lorentz symmetry breaking effects on the relativistic EPR correlations.

The structure of this paper is as follows: In section II, we make a brief introduction to the modified Maxwell theory coupled to gravity that allows us to obtain an effective metric for a curved spacetime under Lorentz symmetry breaking effects; in section III, we introduce the Fermi-Walker transport of spinors and establish a fixed time-like 4-vector in which determine the scenario of the Lorentz symmetry. Then, we calculate the Wigner rotation angle; in section IV, we discuss the effects of the Lorentz violation symmetry on the relativistic EPR correlations and on the Bell inequality; in section V, we present ours conclusions.

\section{geometrical approach}

Due to a lack of a more fundamental theory that, for instance, explain the upper bounds established in experiments for the electron electric dipole moment \cite{revmod,science}, a necessity of investigating the physics beyond the Standard Model has arisen in recent years. A possible way of dealing with a scenario beyond the Standard Model was proposed by Kosteleck\'y and Samuel \cite{extra3} in the string theory, where there exists an extension of the mechanism for the spontaneous symmetry breaking via vector or tensor fields which implies that the Lorentz symmetry is violated. These models that deal with the physics beyond the Standard Model are considered as effective theories and are called as the Standard Model Extension (SME) \cite{colladay-kost}. In this framework, the effective Lagrangian corresponds to the usual Lagrangian of the Standard Model to which is added to the Standard Model operators a Lorentz violation tensor background. The effective Lagrangian is written as an invariant under the Lorentz transformation of coordinates in order to guarantee that the observer independence of physics. However, the physically relevant transformations are those that affect only the dynamical fields of the theory. These changes are called particle transformations, whereas the coordinate transformations (including the background tensor) are called the observer transformations. In Refs. \cite{coll-kost,baeta,bras}, one can find a deep analysis of these concepts. Concerning the experimental searches for the CPT/Lorentz-violation signals, the generality of the SME has provided the basis for many investigations. In the flat spacetime limit, empirical studies include muons \cite{muon}, mesons \cite{meson,meson2}, baryons \cite{barion,barion2}, photons \cite{photon,petrov}, electrons \cite{electron}, neutrinos \cite{neutrino} and the Higgs sector \cite{higgs}. The gravity sector has also been explored in Refs. \cite{gravity,gravity2}. In Ref. \cite{data}, one can find the current limits on the coefficients of the Lorentz symmetry violation. In recent years, Lorentz symmetry breaking effects have been investigated in the hydrogen atom \cite{manoel}, on the Rashba coupling \cite{rash,bb3}, in a quantum ring \cite{bb4}, in Weyl semi-metals \cite{weyl}, in tensor backgrounds \cite{louzada,manoel2}, in the quantum Hall effect \cite{lin2} and geometric quantum phases \cite{belich,belich1,bbs2}.

After the seminal work made by Kosteleck\'y and Mewes \cite{19,20}, a great deal of works \cite{colladay-kost,coll-kost,kost2,gravity,gravity2,curv,curv2,curv3,louzada,cas,bb15} have discussed the extension of the Standard Model in the even sector of SME by the following term: $S=-\frac{1}{4}\int d^{4}x\;K_{abcd}\,F^{ab}\,F^{cd}$. In particular, the tensor $K_{abcd}$ does not violate the CPT-symmetry. It is well-known that the violation of the CPT-symmetry implies that the Lorentz invariance is violated \cite{greenberg}, however, the reverse is not necessarily true. The action $S=-\frac{1}{4}\int d^{4}x\;K_{abcd}\,F^{ab}\,F^{cd}$ breaks the Lorentz symmetry in the sense that the tensor $K_{abcd}$ has a non-null vacuum expectation value. It is worth mentioning that the properties of the tensor $K_{abcd}$ are the same of the Riemann tensor, but it has an additional double-traceless condition. From Refs. \cite{curv3,curv2,curv4}, the tensor $K_{abcd}$ in terms of a traceless and symmetric matrix $\tilde{\kappa}_{ab}$ as $K_{abcd}=\frac{1}{2}\left(\eta_{ac}\,\tilde{\kappa}_{bd}-\eta _{ad}\,\tilde{\kappa}_{bc}+\eta_{bd}\,\tilde{\kappa}_{ac}-\eta_{bc}\tilde{\kappa}_{ad}\right)$. In addition, we can define a normalized parameter 4-vector $\xi^{a}$ that satisfies the conditions: $\xi_{a}\xi^{a}=1$ for the timelike case and $\xi_{a}\xi^{a}=-1$ for the spacelike case; thus, we can decompose the tensor $\tilde{\kappa}_{ab}$ as $\tilde{\kappa}_{ab}=\kappa\left(\xi_{a}\xi_{b}-\frac{\eta_{ab}\,\xi ^{c}\xi_{c}}{4}\right)$, where $\kappa=\frac{4}{3}\tilde{\kappa}^{ab}\,\xi_{a}\,\xi_{b}$. By following Refs. \cite{curv3,curv4}, we consider parameter $\kappa$ to be a spacetime independent parameter, where $0\leq\kappa<2$. Recently, two interesting works \cite{curv2,curv3} have shown that the Lagrange density to the nonbirefringent modified Maxwell theory coupled to gravity can be written in terms of an effective metric tensor $\bar{g}_{\mu\nu}\left(x\right)$ as $\mathcal{L}_{\mathrm{modM}}=-\sqrt{g}\left(1-\frac{1}{2}\,\kappa\,\xi_{\alpha}\,\xi^{\alpha}\right)\,\frac{1}{4}\,F^{\mu\nu}\left(x\right)F^{\rho\sigma}\left(x\right)\,\bar{g}_{\mu\rho}\left(x\right)\,\bar{g}_{\nu\sigma}\left(x\right)$, where this effective metric tensor is given by \cite{curv2,curv3}:
\begin{eqnarray}
\bar{g}_ {\mu\rho}\left(x\right)=g_{\mu\rho}\left(x\right)-\epsilon\,\xi_{\mu}\,\xi_{\rho},
\label{1.4}
\end{eqnarray}
where the parameter $\epsilon$ is defined as $\epsilon=\frac{\kappa}{1+\frac{\kappa}{2}}$ and $\bar{g}^{\mu\nu}\,\bar{g}_{\nu\alpha}=\delta^{\mu}_{\,\,\,\alpha}$. However, the rules of lowering or raising of indices are performed by using the original background metric $g_{\mu\nu}\left(x\right)$ and its inverse $g^{\mu\nu}\left(x\right)$.

In recent years, a geometrical approach to study Lorentz symmetry breaking effects has been proposed based on the Kaluza-Klein theory, by considering the Lorentz violating tensor fields with expectation values along the extra directions \cite{ct,petrov}. Based on this perspective, our proposal is to extend the geometric approximation that gives rise to the effective metric tensor (\ref{1.4}) to a fermionic field by modifying the Minkowski spacetime. In Refs. \cite{bh,bh2,bh3,curv2,curv3} is discussed that the generalized second law of thermodynamics would be violated by the modified Maxwell theory and fermions in the presence of a black hole. However, if the Minkowski spacetime is modified by the effects of a Lorentz symmetry breaking background, therefore there is nothing, at first principles, that prohibit fields and particles described by the Standard Model to feel the anisotropies described by the effective metric tensor (\ref{1.4}). Our proposal is to analyze the spontaneous violation of Lorentz symmetry as an effective theory that can suggest how to move in the direction of a fundamental theory. Therefore, we do not analyze limit situations where singularities appear, such as the event horizon, because in fact we are not sure that these are real situations in a more fundamental theory. Observe that the way of building the effective metric in Eq. (1) results from the decomposition of the tensor $K_{abcd}$ suggested in Refs. \cite{curv2,curv3}, therefore our proposal is to extend the geometric approximation that gives rise to the effective metric tensor in Eq. (1) to a fermionic field by modifying the Minkowski spacetime. By comparing with Refs. \cite{curv4,new}, the background defined by the effective metric in Eq. (1) can yield an analogue effect of the tensor $c_{\mu\nu}$ in the Dirac equation. Both approaches suggest that the violation of the Lorentz symmetry can be view in the Dirac equation through the geometry of the spacetime, but they differ from each other in the way of coupling the terms associated with the violation of the Lorentz symmetry.

Thereby, the Dirac equation can be written in terms of the effective metric tensor $\bar{g}_ {\mu\rho}\left(x\right)$. Despite we do not work with the Dirac equation directly, we deal with the wave function through the WKB approximation \cite{griff} and assume that it is a solution to the Dirac equation in the spacetime background described by the effective metric tensor (\ref{1.4}). Therefore, by assuming that this extension is possible, the aim of this work is to investigate the Lorentz symmetry breaking effects on relativistic EPR correlations in the Minkowski spacetime. By establishing a possible scenario of the Lorentz symmetry violation, we obtain the Wigner rotation angle via the Fermi-Walker transport of spinors and also consider the WKB approximation \cite{griff,anandan} in order that the influence of Lorentz symmetry breaking effects on the relativistic EPR correlations can be discussed.

\section{Fermi-Walker transport of spinors and Wigner rotation}

In this section, we deal with spinors in curvilinear coordinates, then, an appropriate way of working is to use the mathematical approach of spinors in curved spacetime since spinors are defined locally and where spinors transform according to the infinitesimal Lorentz transformations \cite{bd}. Therefore, we need to build a local reference frame for the observers through a noncoordinate basis defined as $\hat{\theta}^{a}=e^{a}_{\,\,\,\mu}\left(x\right)\,dx^{\mu}$, where the components $e^{a}_{\,\,\,\mu}\left(x\right)$ are called \textit{tetrads} and satisfy the relation: $g_{\mu\nu}\left(x\right)=e^{a}_{\,\,\,\mu}\left(x\right)\,e^{b}_{\,\,\,\nu}\left(x\right)\,\eta_{ab}$ \cite{weinberg,bd,naka}, where $\eta_{ab}=\mathrm{diag}(+ - - - )$ is the Minkowski tensor. Besides, the tetrads have an inverse defined as $dx^{\mu}=e^{\mu}_{\,\,\,a}\left(x\right)\,\hat{\theta}^{a}$, and the relations $e^{a}_{\,\,\,\mu}\left(x\right)\,e^{\mu}_{\,\,\,b}\left(x\right)=\delta^{a}_{\,\,\,b}$ and $e^{\mu}_{\,\,\,a}\left(x\right)\,e^{a}_{\,\,\,\nu}\left(x\right)=\delta^{\mu}_{\,\,\,\nu}$ are satisfied. It is interesting to note that if a spinor is transported from a point $x$ of the spacetime to another point $x'$ under the action of external forces, but without torque, the law of transport is given by the Fermi-Walker transport \cite{misner,synge,steph}. For spinors, the Fermi-Walker transport is given by
\begin{eqnarray}
\frac{D}{d\tau}\,e^{a}_{\,\,\,\mu}\left(x\right)=-\frac{1}{c^{2}}\,\left[a_{\mu}\left(x\right)\,U^{\nu}\left(x\right)-U_{\mu}\left(x\right)\,a^{\nu}\left(x\right)\right]\,e^{a}_{\,\,\,\nu}\left(x\right),
\label{2.2}
\end{eqnarray} 
where $\frac{D}{d\tau}$ is the covariant derivative, $U^{\nu}\left(x\right)=\frac{dx^{\nu}}{d\tau}$ is the 4-velocity, $a^{\nu}\left(x\right)=U^{\mu}\left(x\right)\,\nabla_{\mu}\,U^{\nu}\left(x\right)$ is the 4-acceleration ($\nabla_{\mu}$ are the components of the covariant derivative) and $\tau$ is the proper time of a particle. The Fermi-Walker transport was introduced in the quantum mechanical context by Anandan \cite{anandan} through the WKB approximation \cite{griff}, where the wave packet can be Fermi-Walker transported from a initial point $x$ to a final point $x'$ if a particle is moving in an accelerated path, but no torque exists. In the present work, we propose to modify the operator that determines this evolution of the wave packet in order to work with the spinorial algebra. Hence, the modified operator that gives rise to the Fermi-Walker transport of a spinor is given by \cite{bcf2,bcf3}
\begin{eqnarray}
\hat{\Xi}=\hat{P}\exp\left(\frac{i}{4}\int\Omega_{\mu\,a\,b}\left(x\right)\,\,\Sigma^{ab}\,dx^{\mu}\right),
\label{2.3}
\end{eqnarray}
where $\hat{P}$ denotes the path ordering operator, $\Sigma^{ab}=\frac{i}{2}\left[\gamma^{a},\,\gamma^{b}\right]$ is the (spinorial) generator of the Lorentz transformations, $\gamma^{a}$ are the Dirac matrices defined in the Minkowski spacetime \cite{bd,greiner}. The $\gamma^{\mu}$ matrices are related to the $\gamma^{a}$ matrices via $\gamma^{\mu}=e^{\mu}_{\,\,\,a}\left(x\right)\gamma^{a}$ \cite{bd}. Besides, the object $\Omega_{\mu\,a\,b}\left(x\right)$ given in Eq. (\ref{2.3}) is defined as $\Omega_{\mu\,\,\,b}^{\,\,\,a}\left(x\right)=\omega_{\mu\,\,\,b}^{\,\,\,a}\left(x\right)+\tau_{\mu\,\,\,b}^{\,\,a}\left(x\right)$ and it is called as  a connection 1-form or the spin connection \cite{naka,tu2}.  This connection 1-form can be obtained by solving the the Maurer-Cartan structure equations in the absence of torsion $d\hat{\theta}^{a}+\omega^{a}_{\,\,\,b}\wedge\hat{\theta}^{b}=0$ \cite{naka}. Besides, the term $\tau_{\mu\,\,\,b}^{\,\,a}\left(x\right)$ was introduced by Anandan \cite{anandan} and it arises from the action of external forces on the wave function. It is defined as $\tau_{\mu\,\,\,b}^{\,\,a}\left(x\right)=\frac{a^{\nu}\left(x\right)}{c^{2}}\,\left[e_{\,\,\nu}^{a}\left(x\right)\,e_{b\mu}\left(x\right)-e_{\,\,\mu}^{a}\left(x\right)\,e_{b\nu}\left(x\right)\right]$ and it can give rise to quantum effects such as the arising of geometric quantum phases associated with the Thomas precession \cite{anandan}.

In what follows, let us write the Minkowski spacetime in cylindrical coordinates, $ds^{2}=c^{2}\,dt^{2}-d\rho^{2}-\rho^{2}d\varphi^{2}-dz^{2}$, in order to study quantum effects associated with Lorentz symmetry breaking effects in the geometrical picture given by the effective metric tensor (\ref{1.4}). Let us proceed with the choice of the local reference frame for the observers as being $\hat{\theta}^{a}=c\,dt$, $\hat{\theta}^{1}=d\rho$, $\hat{\theta}^{2}=\rho\,d\varphi$ and $\hat{\theta}^{3}=dz$, and  let us consider the normalized parameter 4-vector $\xi_{a}$ to be a time-like 4-vector given by
\begin{eqnarray}
\xi_{a}=\left(1,0,0,0\right).
\label{4.1}
\end{eqnarray}
In this case, we have that the condition $\xi_{\mu}\left(x\right)\,\xi^{\mu}\left(x\right)=\mathrm{const}$ established in Refs. \cite{curv,curv2,curv3} is satisfied, then, the effective metric becomes
\begin{eqnarray}
\bar{ds}^{2}=\left(1-\epsilon\right)\,c^{2}\,dt^{2}-d\rho^{2}-\rho^{2}\,d\varphi^{2}-dz^{2},
\label{4.2}
\end{eqnarray}
and the local reference frame of the observers related to the effective metric given in Eq. (\ref{4.2}) can be defined as
\begin{eqnarray}
\hat{\Theta}^{0}=c\,\sqrt{1-\epsilon}\,\,dt;\,\,\,\hat{\Theta}^{1}=d\rho;\,\,\,\hat{\Theta}^{2}=\rho\,\,d\varphi;\,\,\,\hat{\Theta}^{3}=dz.
\label{4.3}
\end{eqnarray}
By solving the Cartan structure equations in the absence of torsion $d\hat{\Theta}^{a}+\omega^{a}_{\,\,\,b}\wedge\hat{\Theta}^{b}=0$ \cite{naka}, we obtain the  non-null components of the connection 1-form: $\omega_{\varphi\,\,\,1}^{\,\,\,2}\left(x\right)=-\omega_{\varphi\,\,\,2}^{\,\,\,1}\left(x\right)=1$.

Henceforth, let us consider a circular motion with $\rho=\mathrm{const}$ and $z=0$; thus, we can deal with the system in $\left(2+1\right)$ dimensions. The components of the 4-velocity are given by
\begin{eqnarray}
U^{t}\left(x\right)=\frac{c}{\sqrt{1-\epsilon}}\,\cosh\xi,\,\,\,\,\,\,\,U^{\varphi}\left(x\right)=\frac{\sqrt{1-\epsilon}}{\eta\rho}\,c\sinh\xi,
\label{4.5}
\end{eqnarray}
where $\tanh\xi=\frac{v}{c}$, $v=\frac{\eta\rho}{\left(1-\epsilon\right)}\,\,\frac{d\varphi}{dt}$ and he proper time of the particle is  $\tau=\frac{\eta\,\rho}{\sqrt{1-\epsilon}}\,\frac{\Phi}{c\,\sinh\xi}$. Moreover, the only non-null component of the 4-acceleration is $a^{\rho}\left(x\right)=\frac{c^{2}\,\sqrt{1-\epsilon}}{\rho}\,\sinh^{2}\xi$. Hence, the components of the connection 1-form $\tau^{\,\,\,a}_{\mu\,\,\,b}\left(x\right)$ are (with respect to $\left(2+1\right)$ dimensions): $\tau^{\,\,\,0}_{t\,\,\,1}\left(x\right)=\tau^{\,\,\,1}_{t\,\,\,\,0}\left(x\right)=\frac{c\,\left(1-\epsilon\right)}{\rho}\,\sinh^{2}\xi$ and $\tau^{\,\,\,1}_{\varphi\,\,\,2}\left(x\right)=-\tau^{\,\,\,2}_{\varphi\,\,\,1}\left(x\right)=-\eta\sqrt{1-\epsilon}\,\sinh^{2}\xi$, and the non-null components of $\Omega_{\mu\,\,\,b}^{\,\,\,a}\left(x\right)$ are 
\begin{eqnarray}
\Omega^{\,\,\,0}_{t\,\,\,1}\left(x\right)&=&-\Omega^{\,\,\,1}_{t\,\,\,0}\left(x\right)=\frac{c\,\left(1-\epsilon\right)}{\rho}\,\sinh^{2}\xi;\nonumber\\
[-2mm]\label{4.8}\\[-2mm]
\Omega^{\,\,\,1}_{\varphi\,\,\,2}\left(x\right)&=&-\Omega^{\,\,\,2}_{\varphi\,\,\,1}\left(x\right)=-\eta\left(1+\sqrt{1-\epsilon}\,\sinh^{2}\xi\right).\nonumber
\end{eqnarray}

Returning to the operator (\ref{2.3}) that gives rise to the Fermi-Walker transport of a spinor, we have that the Dirac matrices $\gamma^{a}$ are given in $\left(2+1\right)$ dimensions  \cite{matrix,matrix2}: $\gamma^{0}=\sigma^{3}$, $\gamma^{1}=i\sigma^{1}$ and $\gamma^{2}=i\sigma^{2}$, matrices $\sigma^{k}$ are the Pauli matrices and satisfy the relation $\left(\sigma^{i}\,\sigma^{j}+\sigma^{j}\,\sigma^{i}\right)=2\,\delta^{ij}$. Therefore,  the operator (\ref{2.3}) is written as
\begin{eqnarray}
\hat{\Xi}=\exp\left(i\,\frac{\Gamma}{2}\right)=\sum_{n=0}^{\infty}\frac{1}{n!}\,\left(\frac{\Gamma}{2}\right)^{n}=\cos\frac{\beta}{2}+i\,\frac{\Gamma}{\beta}\,\sin\frac{\beta}{2},
\label{4.9}
\end{eqnarray}
where $\Gamma$ and $\Gamma^{2}$ are matrices defined in the form:
\begin{eqnarray}
\Gamma=\left(
\begin{array}{cc}
\eta_{2} & -\eta_{1} \\
\eta_{1} & -\eta_{2} \\
\end{array}\right);\,\,\,\,\,\,\,
\Gamma^{2}=\,\beta^{2}\left(
\begin{array}{cc}
1 & 0 \\
0 & 1 \\
\end{array}\right).
\label{4.10}
\end{eqnarray}

The parameters written in the matrices above are $\eta_{1}=\Phi\,\sinh\xi\cosh\xi$ and $\eta_{2}=\Phi\left[1+\sqrt{1-\epsilon}\,\sinh^{2}\xi\right]$, and the parameter $\beta$ established in Eqs. (\ref{4.9}) and (\ref{4.10}) is   
\begin{eqnarray}
\beta=\pm\Phi\,\sqrt{\left[1+\sqrt{1-\epsilon}\,\sinh^{2}\xi\right]^{2}-\sinh^{2}\xi\,\cosh^{2}\xi}.
\label{4.11}
\end{eqnarray}

The angle $\beta$ given in Eq. (\ref{4.11}) is the angle of the Wigner rotation in the Minkowski spacetime under the influence of Lorentz symmetry breaking effects obtained via the Fermi-Walker transport of a spinor. In the present case, we have established a particular background of the violation of the Lorentz symmetry defined by a fixed time-like 4-vector (\ref{4.1}). Note that the Wigner rotation angle depends on the parameter $\xi$ related to the accelerated motion of the particles, the position of the observers and the Lorentz symmetry violation background. In particular, the effects of the Lorentz symmetry violation background on the Wigner rotation can be viewed through the presence of the parameter $\epsilon$ in Eq. (\ref{4.11}). On the other hand, by taking $\epsilon=0$ in Eq. (\ref{4.11}), the effects of the Lorentz symmetry violation vanish and we recover the Wigner rotation angle in the Minkowski spacetime which is defined by the accelerated motion of the particles, the position of the observers \cite{tu2}.

\section{relativistic EPR correlations and Bell's inequality}

Now, let us deal with the Fermi-Walker transport of spinors in $\left(2+1\right)$ dimensions in the context of the quantum theory. Since classical paths cannot be established in a quantum system, we perform the WKB approximation in order to introduce a geometric approach yielded by the Fermi-Walker transport operator given in Eq. (\ref{4.9}) and study relativistic EPR correlations in a general relativity background. Let us begin by writing the relativistic EPR quantum states in a general relativity background as given in Ref. \cite{tu2}:
\begin{eqnarray}
\left|\psi^{\pm}\right\rangle&=&\frac{1}{\sqrt{2}}\left[\left|p_{+}^{a}\left(x\right),\,\uparrow;\,x\right\rangle\left|p^{a}_{-}\left(x\right),\,\downarrow;\,x\right\rangle\pm\left|p_{+}^{a}\left(x\right),\,\downarrow;\,x\right\rangle\left|p^{a}_{-}\left(x\right),\,\uparrow;\,x\right\rangle\right];\nonumber\\
[-2mm]\label{5.1}\\[-2mm]
\left|\phi^{\pm}\right\rangle&=&\frac{1}{\sqrt{2}}\left[\left|p_{+}^{a}\left(x\right),\,\uparrow;\,x\right\rangle\left|p^{a}_{-}\left(x\right),\,\uparrow;\,x\right\rangle\pm\left|p_{+}^{a}\left(x\right),\,\downarrow;\,x\right\rangle\left|p^{a}_{-}\left(x\right),\,\downarrow;\,x\right\rangle\right],\nonumber
\end{eqnarray}
where $p^{a}\left(x\right)$ is defined in the local reference frame of the observers, $x$ denotes the position of the observers and $\sigma=\uparrow,\,\downarrow$ denotes the spins of the particles. Hence, by definition, each state above transforms under local Lorentz transformations in the spin-half representation at each local reference frame of the observers.

By considering an EPR source and two observers on the plane defined by $\rho=\mathrm{const}$ and $z=0$, where their positions are given by the azimuthal angles $\varphi=0$ and $\varphi=\pm\Phi$, respectively, hence, before the emission of the EPR pair of particles from the source in opposite directions in the circular motion, the initial states of the particle are described by
\begin{eqnarray}
\left|\psi^{-}\right\rangle=\frac{1}{\sqrt{2}}\,\left\{\left|p_{+}^{a}\left(0\right),\uparrow;\,\varphi=0\right\rangle\left|p_{-}^{a}\left(0\right),\downarrow;\,\varphi=0\right\rangle-\left|p_{+}^{a}\left(0\right),\downarrow;\,\varphi=0\right\rangle\left|p_{-}^{a}\left(0\right),\uparrow;\,\varphi=0\right\rangle\right\},
\label{5.2}
\end{eqnarray}
where the 4-momentum of each particle in the local reference frame is given by $p_{\pm}^{a}\left(0\right)=m\,c\,\left(\cosh\xi,\,0,\,0,\,\pm\sinh\xi\right)$, and the spins are parallel to the 1-axis of the local reference frame as in Ref. \cite{tu2}. 

After the emission of the relativistic EPR pair of particles, we consider the WKB approximation in order to describe the Fermi-Walker transport of the quantum state (\ref{5.2}) from the initial point $\varphi=0$ to the final points $\varphi=\pm\Phi$ where the observers are placed. Note that the operator (\ref{4.9}) acts on each spin state of the particle, therefore we label $\hat{\Xi}_{\pm}=\cos\frac{\beta}{2}\pm i\,\frac{\Gamma}{\beta}\,\sin\frac{\beta}{2}$; thus, the operator $\hat{\Xi}_{+}=\cos\frac{\beta}{2}+i\,\frac{\Gamma}{\beta}\,\sin\frac{\beta}{2}$ acts on the spin states of the particle with momentum $p^{a}_{+}$, while the operator $\hat{\Xi}_{-}=\cos\frac{\beta}{2}-i\,\frac{\Gamma}{\beta}\,\sin\frac{\beta}{2}$ acts on the spin states of the particle with momentum $p^{a}_{-}$. Let us consider the observers to be in the rest frame of the particles given in Eq. (\ref{4.3}), then, after applying the Fermi-Walker transport on the quantum state (\ref{5.2}), we obtain the following quantum state in the points $\varphi=\pm\Phi$ where the observers are placed:
\begin{eqnarray}
\left|\zeta\right\rangle=\cos\beta\,\left|\psi^{-}\right\rangle-i\,\cosh\xi\,\sin\beta\,\left|\psi^{+}\right\rangle+i\,\sinh\xi\,\sin\beta\,\left|\phi^{+}\right\rangle.
\label{5.3}
\end{eqnarray}

Observe that the Fermi-Walker transport operator (\ref{4.9}) acts on the spinors, that is, on the spin states, thus, the quantum state of the correlated particles obtained in Eq. (\ref{5.3}) is analogous to the quantum states obtained in Ref. \cite{bcf} via infinitesimal Lorentz transformations (applied at each point of the spacetime from $\varphi=0$ to $\varphi=\pm\Phi$) in the sense that the spins of the relativistic correlated particles in the initial EPR correlation undergo a precession that depends on the angle $\beta$, while the 4-momentum of the particles remains unchanged in the local reference frame of the observers \cite{anandan}. However, we can see that the final state of the relativistic correlated particles is given by the superposition of the states $\left|\psi^{-}\right\rangle$, $\left|\psi^{+}\right\rangle$ and $\left|\phi^{+}\right\rangle$ and depends on the angle of the Wigner rotation $\beta$ and the parameter $\xi$, which differs from the state obtained in Refs. \cite{bcf}. Moreover, no spurious effects from arbitrary rotations of the local axis exist, since the Fermi-Walker reference frame is built in such a way that the local axis do not rotate.

As pointed out in Refs. \cite{tu2,bcf}, the states of the relativistic correlated particles after undergoing the Wigner rotation suggests that the initial spin anticorrelations given in Eq. (\ref{5.2}) are broken if the observers measure the spin in the direction of the 1-axis. Actually, the quantum state (\ref{5.3}) shows that the direction of the spin measurements in which the observers can make at the points $\varphi=\pm\Phi$ must be rotated about the 3-axis of the local reference frame of observers through the angles $\mp\beta$, respectively. This means that, by knowing the relativistic effects that stem from the acceleration of the particles and the background of the Lorentz symmetry violation, we can recover the initial spin anticorrelations given in Eq. (\ref{5.2}) by rotating the spin axis of measurement through the angles $\mp\beta$ about the 3-axis of the local reference frame of the observers at $\varphi=\pm\Phi$.

Let us analyse the violation of the Bell inequality in this system. Since the particles are moving in a circular motion on the plane defined by $\rho=\mathrm{const}$ and $z=0$, then, suppose that the first observer is placed in $\varphi=+\Phi$ and measure the component of the spin through the observables $\hat{a}$ and $\hat{a}'$, while the second observer is placed in $\varphi=-\Phi$ and measure the component of the spin through the observables $\hat{b}$ and $\hat{b}'$, where these operators are defined as $\hat{a}=\frac{\sigma^{1}+\sigma^{3}}{\sqrt{2}}$, $\hat{a}'=\frac{\sigma^{3}-\sigma^{1}}{\sqrt{2}}$, $\hat{b}=\sigma^{3}$ and $\hat{b}'=\sigma^{1}$. By following Ref. \cite{nc}, the maximum violation of the Bell inequality can be given by $\left|\left\langle \hat{a}\,\hat{b}\right\rangle+\left\langle \hat{a}'\,\hat{b}\right\rangle+\left\langle \hat{a}\,\hat{b}'\right\rangle-\left\langle \hat{a}'\,\hat{b}'\right\rangle\right|\leq2\,\sqrt{2}$. Thereby, by taking the quantum state given in Eq. (\ref{5.3}), we obtain
\begin{eqnarray}
\left|\left\langle \hat{a}\,\hat{b}\right\rangle+\left\langle \hat{a}'\,\hat{b}\right\rangle+\left\langle \hat{a}\,\hat{b}'\right\rangle-\left\langle \hat{a}'\,\hat{b}'\right\rangle\right|=2\,\sqrt{2}\,\left|\cosh^{2}\xi\,\sin^{2}\beta-1\right|.
\label{5.6}
\end{eqnarray}

We can see in Eq. (\ref{5.6}) that the violation of the Bell inequality depends on the angle of the Wigner rotation $\beta$ given in Eq. (\ref{4.11}) and the parameter $\xi$. Observe that the violation of the Bell inequality obtained from the quantum state of the correlated particles (\ref{5.3}), given by the Fermi-Walker transport of spinors, is different with respect to that obtained in Ref. \cite{bcf}. This dependence on the angle $\beta$ and the parameter $\xi$ means that the degree of violation of the Bell inequality decreases by making the spin measurements on the 1-axis of the local reference frame of the observers at $\varphi=\pm\Phi$. This result stems from the relativistic effects of the Lorentz symmetry violation background and the accelerated motion of the correlated particles in the plane defined by $z=0$ and $\rho=\mathrm{const}$. However, by rotating the the spin axes of measurements through the angles $\mp\beta$ about the 3-axis of the local reference frame of the observers at $\varphi=\pm\Phi$, we can recover both the initial spin anticorrelations and the maximum violation of the Bell inequality.

\section{conclusions}

From the description of fermions in a curved spacetime in the presence of Lorentz symmetry breaking effects, we have established a possible scenario of the Lorentz symmetry violation in the Minkowski spacetime and obtained the Wigner rotation angle by using the Fermi-Walker transport of spinors. We have seen that the angle of the Wigner rotation is determined by the accelerated motion of the particle, the position of the observers and the Lorentz symmetry violation background. We have also discussed a possible application of this study in relativistic EPR correlations by using the WKB approximation, where a precession of the spins of the correlated particles and the decrease in the degree of the violation of the Bell inequality are observed by making the spin measurements on the 1-axis in local frame of the observers. However, we have seen that by knowing the relativistic effects, we can recover the initial spin anticorrelations and the maximum violation of the Bell inequality by correcting the direction of the spin axis in terms of the Wigner rotation angle. Although experiments with relativistic spin-$1/2$ particle are very hard to perform, the present study suggests that experiments with relativistic EPR correlations could be a possible way of testing the violation of the Lorentz symmetry in the Minkowski spacetime as well as in a scenario of general relativity.

\acknowledgments{The authors would like to thank CNPq (Conselho Nacional de Desenvolvimento Cient\'ifico e Tecnol\'ogico - Brazil) for financial support.}

\end{document}